\documentclass[sn-mathphys,Numbered]{modified-jnl}

\usepackage{graphicx}%
\usepackage{multirow}%
\usepackage{mathrsfs}%
\usepackage[title]{appendix}%
\usepackage{xcolor}%
\usepackage{textcomp}%
\usepackage{manyfoot}%
\usepackage{booktabs}%
\usepackage{algorithm}%
\usepackage{algorithmicx}%
\usepackage{algpseudocode}%
\usepackage{listings}%

\newcommand{\pt}{\ensuremath{p_\text{T}}}
\newcommand{\met}{\ensuremath{p_{\text{T}}^{\text{miss}}}}
\newcommand{\xaod}{\texttt{xAOD}}

\begin{document}

\title{A flexible and efficient approach to missing transverse momentum reconstruction}


\author*[1]{\fnm{William} \sur{Balunas}}\email{bill.balunas@cern.ch}
\author[2]{\fnm{Donatella} \sur{Cavalli}}
\author*[3]{\fnm{Teng Jian} \sur{Khoo}}\email{teng.jian.khoo@cern.ch}
\author[4]{\fnm{Matthew} \sur{Klein}}
\author[5]{\fnm{Peter} \sur{Loch}}
\author[6]{\fnm{Federica} \sur{Piazza}}
\author[2]{\fnm{Caterina} \sur{Pizio}}
\author[2]{\fnm{Silvia} \sur{Resconi}}
\author[7]{\fnm{Douglas} \sur{Schaefer}}
\author[8]{\fnm{Russell} \sur{Smith}}
\author[1]{\fnm{Sarah} \sur{Williams}}

\affil[1]{\orgdiv{Cavendish Laboratory}, \orgname{University of Cambridge}, \orgaddress{\city{Cambridge}, \country{United Kingdom}}}

\affil[2]{\orgdiv{INFN Sezione di Milano \& Dipartimento di Fisica}, \orgname{Universit\`a di Milano}, \orgaddress{\city{Milano}, \country{Italy}}}

\affil[3]{\orgdiv{Institut f\"ur Physik}, \orgname{Humboldt-Universit\"at zu Berlin}, \orgaddress{\city{Berlin}, \country{Germany}}}

\affil[4]{\orgdiv{Physics Department}, \orgname{Southern Methodist University}, \orgaddress{\city{Dallas}, \state{TX}, \country{United States of America}}}

\affil[5]{\orgdiv{Physics Department}, \orgname{University of Arizona}, \orgaddress{\city{Tucson}, \state{AZ}, \country{United States of America}}}

\affil[6]{\orgdiv{Institute for Fundamental Science}, \orgname{University of Oregon}, \orgaddress{\city{Eugene}, \state{OR}, \country{United States of America}}}

\affil[7]{\orgdiv{Enrico Fermi Institute}, \orgname{University of Chicago}, \orgaddress{\city{Chicago}, \state{IL}, \country{United States of America}}}

\affil[8]{\orgdiv{Nevis Laboratory}, \orgname{Columbia University}, \orgaddress{\city{Irvington}, \state{NY}, \country{United States of America}}}


\abstract{Missing transverse momentum is a crucial observable for physics at hadron colliders, being the only constraint on the kinematics of ``invisible'' objects such as neutrinos and hypothetical dark matter particles. Computing missing transverse momentum at the highest possible precision, particularly in experiments at the energy frontier, can be a challenging procedure due to ambiguities in the distribution of energy and momentum between many reconstructed particle candidates. This paper describes a novel solution for efficiently encoding information required for the computation of missing transverse momentum given arbitrary selection criteria for the constituent reconstructed objects. Pileup suppression using information from both the calorimeter and the inner detector is an integral component of the reconstruction procedure. Energy calibration and systematic variations are naturally supported. Following this strategy, the ATLAS Collaboration has been able to optimise the use of missing transverse momentum in diverse analyses throughout Runs 2 and 3 of the Large Hadron Collider and for future analyses.}




\maketitle

\section{Introduction}
\label{sec:intro}

Despite the highly developed state of particle detector design in the era of the Large Hadron Collider (LHC)~\cite{Evans:2008zzb}, there exist particles which even the most sensitive instruments cannot reliably detect.
Neutrinos and similar particles that are practically undetectable (or ``invisible'') are the signifying features of numerous processes of interest, including Standard Model electroweak physics and more hypothetical Beyond-the-Standard Model (BSM) processes such as production of supersymmetric particles or dark matter.
At hadron colliders, the main constraint on invisible particle kinematics in a given event is the missing transverse momentum, the negative vector sum of the transverse momenta of all objects associated with the recorded event.
This two-dimensional vector is denoted $\vec{p}_{\text{T}}^{\text{miss}}$, with its magnitude as \met, $E_{\text{T}}^{\text{miss}}$, or where mathematical notation is infeasible (e.g. in software), ``MET''\footnote{This paper uses the notation \met as this represents a momentum rather than an energy, but ``MET'' or ``Missing $E_{\text{T}}$'' remains ubiquitous for historical reasons. These all denote the same variable.}.
This paper describes a novel approach for reconstructing \met\ that preserves a great deal of flexibility to meet the diverse needs of physics analyses.
It was adopted by the ATLAS experiment~\cite{PERF-2007-01} within the context of its software suite~\cite{ATL-SOFT-PUB-2021-001} for the simulation, reconstruction, and analysis of collision data and has been in use since LHC Run 2.

Due to its status as an inclusive event observable, the reconstruction of \met\ requires the imposition of a global event description: the selection of all objects associated with the hard interaction, and identification or classification of these objects in order to perform calibration.
Hitherto, the definition of this event description has been effectively static~\cite{PERF-2014-04}, fixed at the point when the experimental data are reconstructed, which due to CPU constraints can only performed occasionally.
While this makes the \met\ computation straightforward to implement, this static definition is a limitation in several ways:
\begin{itemize}
\item Particle identities are fixed in the \met\ on long timescales and shared across all use cases, limiting options for optimising these for a given analysis.
\item Particle selection must be made on the basis of calibrations derived at the time of reconstruction, which may be inconsistent with updated calibrations used in the final analysis.
\item Systematic uncertainties cannot be fully accounted for in the particle selection, particularly where these uncertainties can affect reconstructed particle identities.
\end{itemize}
To partially address these issues, the ATLAS \met\ event data model (EDM) has historically incorporated a scheme for book-keeping, recording exactly which reconstructed objects were used in the \met\ calculation and any modifications to the kinematics of these objects applied in the process.

In this paper, an improved \met\ EDM is described, which compactly records information necessary to freely recompute the \met\ given an arbitrary event description and any necessary four-momentum corrections applied to the constituent objects.
This is accomplished by encoding information needed for resolving signal overlaps between different selected objects, using the constituents of hadronic jets~\cite{Salam:2010nqg} as a basis.
The new ``dynamic'' EDM effectively addresses the limitations mentioned above, facilitating efficient optimisation of object selections and \met computations, as well as fully consistent treatment of systematic uncertainties.
An intuitive user interface is critical to ensuring that customisation of the \met\ calculation can be performed by a wide user base.

The outline of the paper is as follows.
Motivations for the dynamic \met\ EDM are given in the context of the ATLAS software and constraints on \met\ reconstruction in Section~\ref{sec:motivation}.
A full description of the EDM follows in Section~\ref{sec:edm}.
Then, Section~\ref{sec:reconstruction} explains the algorithmic implementation of the \met\ reconstruction, showing how information is compiled into the EDM encoding.
The user interface for analysis is detailed in Section~\ref{sec:interface}.
Section~\ref{sec:performance} demonstrates the performance gains in CPU and disk usage from the new approach.
Adaptation of the design to address recent and future challenges in the LHC computing environment are described in Section~\ref{sec:evolution}.
Finally, conclusions are presented in Section~\ref{sec:conclusion}.

\section{Motivation}
\label{sec:motivation}

The design of the ATLAS \met\ reconstruction is foremost specified by the need for a compact data structure that permits a flexible reconfiguration of the \met\ calculation, while ensuring that the calculated observable is robust against reconstruction errors.
While full flexibility could be achieved by retaining in analysis data formats all objects needed for the \met\ computation, the disk cost would be prohibitive due to the extremely high multiplicity of low-energy signals.
Furthermore, \met\ reconstruction necessitates non-trivial operations to disambiguate the calorimeter and tracker signals that may be shared between multiple reconstructed objects.
This makes the procedure more complex than a simple summation over selected particles, and implies the need for a supporting infrastructure to facilitate the \met\ computation in the end stages of analysis event processing, when the final object calibrations are available.
Below, a detailed description of the constraints on ATLAS \met\ reconstruction are given, which determine the specifications for the \met\ EDM and associated analysis tools.
Despite the attention paid here to the specific case of ATLAS, these considerations can be taken as representative of a typical general purpose particle detector operating at a hadron or lepton collider.

\subsection{Detector structure}

The ATLAS detector possesses a concentric cylindrical structure optimised both for particle identification and energy/momentum measurements~\cite{PERF-2007-01}.
The inner detector (ID), immersed in a solenoidal magnetic field, provides precision reconstruction of charged-particle trajectories.
Electromagnetic and hadronic showers are captured in the calorimeter system, with coverage close to 4$\pi$ in solid angle.
The calorimeters possess longitudinal as well as transverse segmentation, to capture shower development in depth.
These components are surrounded by a muon spectrometer (MS) integrated with ATLAS's eponymous toroidal magnet system.
The sub-detectors of ATLAS vary in their fiducial acceptance, with the ID coverage being similar to but more limited than that of the MS, while the calorimeter coverage extends significantly further into the forward region than the other components.

\subsection{Overview of ATLAS event reconstruction}

Reconstruction of analysis objects in ATLAS from the digital detector output, or from analogous simulated inputs, takes place in several steps.
First, \textit{basic constituents} are constructed from the raw digital signals:
\begin{itemize}
	\item Hits in the inner detector are fitted to produce a set of \textbf{tracks} describing the trajectories of charged particles~\cite{Cornelissen:2007vba}.
	\item Calorimeter cell energies are determined and calibrated based on the sampled calorimeter pulse shapes to the scale of electron and photon showers as measured in test beams (electromagnetic scale).
	      Cells are grouped into noise-suppressed \textbf{topological clusters} (topoclusters)~\cite{PERF-2014-07}.
	      The resulting clusters may subsequently be calibrated to correct their energies to match the scale of hadrons using cell-level weights (hadronic scale).
	\item \textbf{Track segments} are formed from hits in the muon spectrometer, which may further be combined into muon spectrometer standalone tracks~\cite{MUON-2018-03}.
\end{itemize}
Subsequently, these basic constituents are combined to form particle candidates or hadronic jets.
Of crucial importance is the fact that most higher-level reconstruction operations of this nature run independently, such that the outputs of different particle identification (PID) algorithms in most cases do not influence one another.
The following \textit{objects} are reconstructed:
\begin{itemize}
	\item \textbf{Electron candidates} are identified based on the presence of narrow showers in the electromagnetic calorimeter~\cite{EGAM-2018-01}.
	      At least one nearby track is associated to the electron candidate and used to refine the energy/momentum measurement.
	      Quality criteria based on shower shape and track properties are defined to provide a balance between high efficiency and a low misidentification rate of fake electron candidates.
	\item \textbf{Photon candidates} are identified similarly to electron candidates, but no track is required; nearby tracks may be used to improve the four-momentum measurement and PID under the assumption that the photon has undergone a conversion in the inner detector material~\cite{EGAM-2018-01}.
	\item \textbf{Muon candidates} are identified using inner detector and/or muon spectrometer tracks~\cite{MUON-2018-03}.
	      The most precise reconstruction and PID is achieved when extrapolated ID tracks can be matched to a MS track, however muon candidates may also be formed using a more limited set of hits in either system, to improve coverage and reconstruction efficiency.
	      Calorimeter cells along the muon trajectory are identified and used to improve estimates of the energy deposited in the calorimeter by the muon.
	\item A particle flow algorithm is run over the tracks and topoclusters, to extract a better measurement of the kinematics of charged hadrons and permit suppression of charged hadron pileup contributions to hadronic measurements~\cite{PERF-2015-09}.
	      Muon and electron candidate tracks are excluded from the particle flow algorithm, as the energy subtraction is optimised for pion showers.
	      The outputs of the particle flow algorithm are termed \textbf{particle flow objects} (PFOs).
	\item \textbf{Jets} are reconstructed using sequential clustering algorithms (usually anti-\(k_t\)~\cite{antikt}, as implemented in \textsc{FastJet}~\cite{Fastjet}) from topoclusters calibrated at either the hadronic or electromagnetic scales, or charged and neutral particle flow objects~\cite{PERF-2015-09,JETM-2018-05}.
	      The inputs provided to the jet clustering algorithm are henceforth designated \textit{jet constituents}.
	      Tracks in the catchment area of each jet are matched to the jet and used for calibration and pileup suppression.
	      A sequence of calibrations is applied to subtract pileup contributions, match the average jet scale to that of simulated hadrons, compensate for response differences due to flavour and shower development and finally to correct differences between the response determined in simulation and in data~\cite{JETM-2018-05}.
	\item Hadronically decaying \textbf{tau lepton candidates} are seeded from jets reconstructed from hadronic scale topoclusters. Particle flow techniques are used to refine the tau energy scale calibration and the description of the tau decay and shower~\cite{PERF-2014-06}. Multivariate classifiers are used to discriminate hadronic taus from parton-initiated jets and electrons~\cite{ATLAS-CONF-2017-029}.
\end{itemize}
Computation of \met\ requires the selection of objects that represent a coherent description of the event, with basic constituents used no more than once in the summation.
Hence objects that share tracks or clusters must undergo an overlap removal procedure.
Fully reconstructed and calibrated objects account only for a limited fraction of the total momentum from the hard scattering process, and therefore the remainder must be estimated from basic constituents unassociated to the selected objects.

\subsection{Analysis requirements on \met\ reconstruction}

Analyses using ATLAS data vary widely in their requirements, with the target final state and dominant background processes heavily influencing exactly which objects are selected for analysis.
To correctly handle systematic uncertainties and to ensure a coherent description of each event, the objects from which the \met\ is calculated must be consistent with the objects defining the event selection and other event reconstruction procedures.
This leads to a set of specific requirements on the \met\ EDM.

First and foremost, the ability to freely choose exactly which leptons, photons, and jets are accepted is important for analysis optimisation.
Optimisation of an analysis requires the ability to repeat the analysis procedure in approximately a day, if not faster, whereas data samples for analysis are reconstructed afresh at most a handful of times in a year.
The use of additional intermediate dataset formats permits a higher frequency for reconstruction operations to be repeated, but is still limited to a timescale of weeks or (more often) months for update of configurations.
This significantly disadvantages the use of static \met\ calculations, as even the provision of dozens of \met\ configurations fails to satisfy highly optimised analyses, and comes at a substantial cost of disk space and CPU.
A non-negligible consideration is the occasional need to fix bugs in the \met\ calculation itself as well as optimisation of the \met\ computation.

PID decisions and four-momentum calibrations are applied at the time of reconstruction, but are typically re-applied during event processing for analysis, taking advantage of refinements derived during the course of data-taking.
For this reason, a computation of \met\ solely based on information available in the initial reconstruction procedure will not correspond to the calibrations and PID used in analysis event selection.
This implies that the \met\ calculation must be able to be updated with the final calibrations and input selection at analysis time.
Even in the case where input selection and calibration could be frozen in advance, the application of systematic uncertainties on object four-vectors taking into account the overlap removal and other corrections applied during \met\ reconstruction would imply that additional information about the contribution of each object to the \met\ sums must be recorded.
Furthermore, the correct handling of systematic uncertainties requires mutability of object selection, as object identities may themselves be subject to uncertainties.

\subsection{Summary of demands on EDM and analysis tools}
\label{sec:motivation:demands}

To address the requirements previously described, the following were determined to be necessary features that must be provided by the \met\ EDM:
\begin{enumerate}
	\item The EDM should record the full space of \textit{possible overlaps} between reconstructed objects rather than forming a simple kinematic sum after overlap removal.
	\item The EDM and supporting tools must permit users to specify lists of selected leptons and photons defining priorities for each class of object to be retained during overlap removal.
	\item The EDM must be capable of determining the momentum sum of basic constituents not associated to selected objects.
	\item The EDM and reconstruction procedure must support differing signal bases for jet reconstruction: topoclusters and tracks or particle flow objects.
\end{enumerate}

In the next section, an implementation of an EDM that satisfies the criteria above is detailed.

\section{Event data model}
\label{sec:edm}

A C++ implementation of this dynamic \met\ EDM is carried out in the context of the \xaod\ EDM devised by ATLAS for LHC Run 2~\cite{xAOD} (2015--2018).
As context for the \met\ EDM description, the \xaod\ data structure is first briefly described, together with relevant fundamental elements of the ATLAS EDM.

\subsection{Overview of the \xaod\ EDM}

The \xaod\ structures data in a tree, using the \texttt{TTree} class from the \texttt{ROOT} framework~\cite{ROOT}, but emulates the organisation of these data into objects (\texttt{AuxElement}), which may be grouped into \emph{containers} (\texttt{AuxVectorData}).
Classes deriving from \texttt{AuxElement} provide the object-oriented interface, whereas the underlying data are stored in an auxiliary store comprising a set of \texttt{std::vector} data members.
For the $i^\mathrm{th}$ element of a container, the corresponding data is held in the $i^\mathrm{th}$ elements of each \texttt{std::vector} in the auxiliary container.
Data on \xaod\ objects may be static (i.e. defined explicitly in the auxiliary container) or dynamic.
Dynamic data may be applied as a decoration, augmenting the information content even of immutable objects.

Containers serve not only as the receptacle for data content, but also as the vessel for information transfer between algorithmic components, being recorded in a \emph{store} with access given by a corresponding \emph{key}.
To provide persistent references to individual objects, the \texttt{ElementLink} construct is used, which identifies a given object by the key of its container and the index of the object within the container.
Concretely, the \texttt{ElementLink} is a template class, taking the target container type as template argument.

A common base class is shared by all reconstructed \xaod\ objects possessing four-momenta: the \texttt{IParticle} class, which itself derives from \texttt{AuxElement}.
Apart from serving as a base type, the \texttt{IParticle} interface chiefly provides access to the basic four-vector, and a check of the type of the object via a C++ \texttt{enum} named \texttt{ObjectType}.

The derived classes of \texttt{IParticle} with relevance to the \met\ EDM at the time of writing include:
\begin{itemize}
\item \texttt{CaloCluster},
\item \texttt{TrackParticle},
\item \texttt{PFO} (superseded by \texttt{FlowElement} for Run 3),
\item \texttt{TruthParticle},
\item \texttt{Jet},
\item \texttt{Electron},
\item \texttt{Photon},
\item \texttt{Muon},
\item \texttt{TauJet}.
\end{itemize}
It should be noted that reconstructed interaction or decay vertices and the \met\ kinematic information are not represented by \texttt{IParticles}.

\subsection{Description of the \met\ data classes}

Two sets of \texttt{xAOD} objects are defined to hold the \met\ data:
\begin{itemize}
	\item \texttt{MissingET} object --- This contains kinematic information with grouping for related terms.
	\item \texttt{MissingETAssociationMap} --- This is a representation of dynamic calculation, including signal-base four-momentum sums for additive/subtractive overlap removal.
\end{itemize}
Each set of classes is described in the following sections.

\subsubsection{\texttt{MissingET}}

The ultimate goal of the \met\ reconstruction is to provide the \met\ two-vector, corresponding to the best estimate of the total vectorial transverse momentum carried by particles produced in the primary interaction which are non-interacting and stable on detector scales.
As such, the main EDM object with direct physical significance is a representation of this two-vector, augmented with supporting information including identification of the contributing objects in the form of a bitmask (``source'' tag) and the scalar transverse momentum sum, which captures information important to \met\ performance characterisation.
The interface object is called the \texttt{MissingET}, which is ordinarily held in a \texttt{MissingETContainer}.

\begin{figure*}[tbp]
  \centering
  \includegraphics[width=0.9\textwidth]{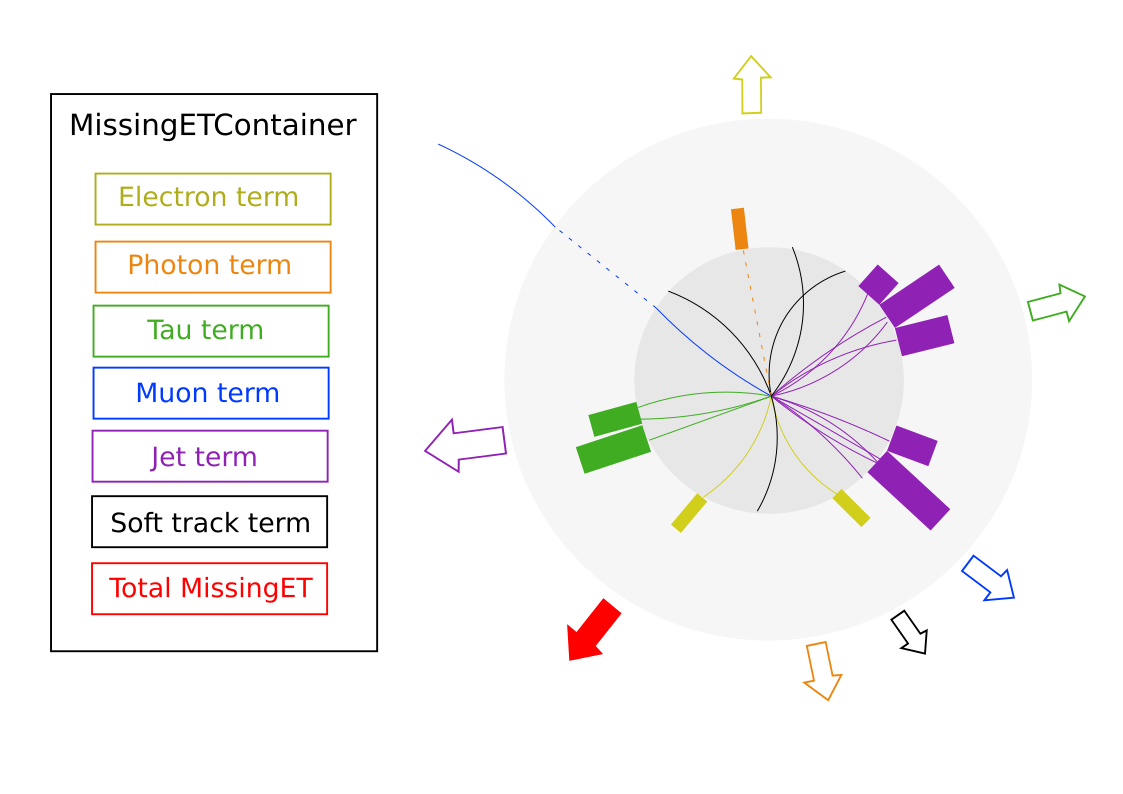}
  \caption{
    \label{fig:metSketch}
    Illustration of the high-level structure of the \texttt{MissingETContainer} EDM and the typical basic constituents (tracks and calorimeter energy clusters) corresponding to each element.
    The ``soft track term'' in this example represents the contribution from tracks not associated to any particle candidate.
    The arrows represent the overall direction of each term.
  }
\end{figure*}

For additional information, the contributions to the total \met\ from different types of particle candidates are encoded in distinct \texttt{MissingET} objects, held in the same \texttt{MissingETContainer} as the total \met.
These are distinguished and retrieved from the container primarily by name, via a fast hash comparison, but can also be extracted by the ``source'' tag.
A qualitative sketch of the high-level structure is shown in Figure~\ref{fig:metSketch}.

\subsubsection{\texttt{MissingETAssociationMap}}
\label{sec:metassoc}

To satisfy the demands listed in Section~\ref{sec:motivation:demands}, a compact representation is needed of the possible combinations of distinct objects whose transverse momenta should be summed to compute the total \met\ two-vector.
The \texttt{MissingETAssociationMap} encodes this information efficiently, permitting overlap removal of the contributing energy/momentum measurements for an arbitrary choice of quality criteria to be applied to the lepton/photon candidates used in the calculation.
This overlap removal is able to be carried out precisely, down to the level of the individual basic constituents (tracks/clusters/PFOs) used to reconstruct the contributing physics objects.

\begin{figure*}[tbp]
  \centering
  \includegraphics[width=0.9\textwidth]{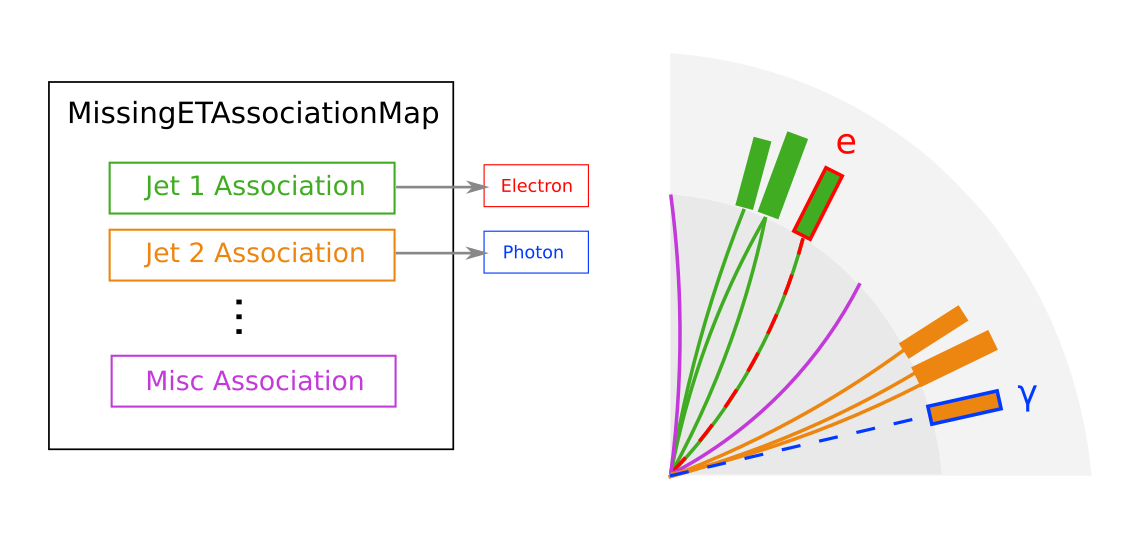}
  \caption{
    \label{fig:metAssocSketch}
    Illustration of the high-level structure of the \texttt{MissingETAssociationMap} EDM and the typical detector signals (tracks and calorimeter energy clusters) corresponding to each element.
  }
\end{figure*}

To condense the necessary information for the overlap removal procedure, jet constituents are used as a basis on which to represent the energy/momentum contributions of every lepton/photon candidate to the total energy measured in each collision event.
Leptons and photons are then matched to jets on the basis of shared basic constituents, limiting the search space needed to determine signal overlaps between leptons and photons.
The \texttt{MissingETAssociationMap} can thus be constructed from a set of individual \texttt{MissingETAssociation} objects.
For an event with $N_\mathrm{jet}$ reconstructed particle-jets, $N_\mathrm{jet}+1$ \texttt{MissingETAssociation} objects are required; one per jet, and one ``miscellaneous'' association, recording the signal contributions that were not matched to any jet.
These exist because jet clustering algorithms such as anti-$k_t$ include an energy or momentum cutoff below which a clustered object is not considered a jet.
A simple example is illustrated in Figure~\ref{fig:metAssocSketch}.

Concretely, each association holds the following information:
\begin{description}
\item[\texttt{jetLink}] An \texttt{ElementLink} to the association's \textit{reference jet}, whose constituents make up the basis for all computations with this association object. For the miscellaneous association, an invalid link is recorded.
\item[\texttt{isMisc}] A \texttt{bool} indicating if this association object is the miscellaneous association.
\item[\texttt{objectLinks}] A \texttt{std::vector<ElementLink>}, identifying the leptons/photons sharing basic constituents with the reference jet. A lepton/photon may share constituents with multiple reference jets, and hence be represented in multiple association objects.
\item[\texttt{overlapIndices}] A \texttt{std::vector<std::vector<size\_t> >}, holding a sparse representation of the overlaps between leptons/photons linked in the \texttt{objectLinks} vector. For each element in the \texttt{objectLinks}, the indices to any other elements of \texttt{objectLinks} that share constituents are stored, with the representation being symmetric.
\item[\texttt{calpx/py/pz/e/sumpt}] Five \texttt{std::vector<float>} members recording the four-momentum and scalar \pt sum of each \textit{basis group of constituents needed to perform overlap removal on the reference jet and associated leptons/photons}, as described below.
\item[\texttt{calkey}] A \texttt{std::vector<bitmask\_t>}\footnote{Typedef of \texttt{unsigned long long}.}, index parallel with the \texttt{calpx/py/pz/e/sumpt} vectors, encoding the associations of the constituent basis groups to the objects referred to by \texttt{objectLinks}.
\end{description}

To mitigate sensitivity to pileup, an alternate representation of the event is formed using only information from charged particles matched to the nominal hard-scatter primary vertex (commonly identified based on associated track momenta), localised by ghost-associating~\cite{Cacciari:2007fd,Cacciari:2008gn} inner detector tracks to the jets.
This is computed in the same way as the calorimeter-based/inclusive representation of the event, but requires additional information to be stored in each \texttt{MissingETAssociation} object:
\begin{description}
\item[\texttt{trkpx/py/pz/e/sumpt/trkkey}] Analogous information to the \texttt{calpx, ...} vectors, computed from selected tracks ghost-associated to the reference jet.
\item[\texttt{jettrkpx/py/pz/e/sumpt}] The four-momentum and scalar \pt sum of the selected tracks ghost-associated to the reference jet.
\item[\texttt{overlapTypes}] A \texttt{std::vector<std::vector<unsigned char> >}, with entries corresponding to the indices in \texttt{overlapIndices}, for which each element functions as a bitmask identifying the types of basic constituents that were found to be shared between the overlapping objects. This representation is primarily used to indicate whether the overlaps are between contributing charged particle tracks or if calorimeter energy clusters are also shared.
\end{description}

Not all basic constituents in a given event are used to reconstruct higher level particle candidates.
The \met\ corresponding to only these leftover basic constituents is recorded in a separate \texttt{MissingET} object, referred to as the ``core soft term'' for the event, which is effectively inert as far as overlap removal is concerned.

\subsubsection*{Overlap removal encoding}
\label{sec:edm:ORencoding}

Within each \texttt{MissingETAssociation}, the possible overlaps between combinations of selected objects are decomposed according to the following process, in order to generate the \textit{overlap removal basis}:
\begin{enumerate}
\item For each of the $N_\mathrm{obj}$ lepton/photons matched to this association, extract the list of basic constituents amongst the reference jet's constituents, and generate a constituent-to-particle-candidate map. Sort the particle candidates by any stable ordering principle.
\item Assign each particle candidate a boolean flag indicating whether this object was selected for the \met\ computation. The state of these flags can be represented by a binary string of length $N_\mathrm{obj}$, with the maximal value of this string being $2^{N_\mathrm{obj}}-1$. The particle candidate with index $i$ will be represented by the bit corresponding to $2^i$.
\item Let $\vec{C}_j$ be a 5-dimensional vector with elements representing the four-momentum coordinates $p_x,p_y,p_z,E$ and the $\pt$ of the jet constituent $j$. Let $\mathcal{M}$ be a map of binary strings $o$ to 5-d vectors $V_o$ where $o \in [0,2^{N_\mathrm{obj}}-1]$.
\item For each jet constituent, check whether it is associated to any particle candidates. If not, continue. If it is, then define $o_j = \sum_{i \in \text{matched particle candidates}} 2^i$, and add $C_j$ to $\mathcal{M}(o_j)$. Record the \texttt{overlapIndices} and set the calorimeter/inclusive bit for \texttt{overlapType} for each corresponding particle candidate $i$.
\item For each $o,V_o$ pair in $\mathcal{M}$ with non-zero $V_o$, add an entry to \texttt{calkey} and \texttt{calpx/py/pz/e/sumpt}, filling in the corresponding components respectively.
\end{enumerate}
The process is repeated substituting ghost-associated hard-scatter tracks for jet constituents in order to fill the \texttt{trkkey/px/py/pz/e/sumpt} vectors.

\section{Reconstruction implementation}
\label{sec:reconstruction}

The central goal of the ``reconstruction'' step is to construct the objects needed at analysis level to compute the \met\ with the desired flexibility.
These consist of the \texttt{MissingETAssociationMap} encoding all possible overlaps between hard objects (jets, leptons and photons satisfying particle identification criteria), as well as a representation of the ``core soft term'' (the constituents not associated to any hard object) in the form of a \texttt{MissingETContainer}.
Finally, these objects can be used to build the final \texttt{MissingET} object according to any given analysis-level object selection definitions.

\subsection{Building the association map}
\label{sec:recomap}

The primary challenge in defining these objects is that the jets, leptons, and photons are created from different detector signals.
Depending on the type of particle, these can include tracks, PFOs, topoclusters, or other specialized detector-level objects.
Therefore, a method is required for associating jet constituents and tracks with leptons and photons that were not necessarily built from them.
This association of constituents to these hard objects is performed by a tool known as a \texttt{MissingETAssociator}.
This is a base class which has a different specialization for each type of hard object, implementing the appropriate method for associating the constituents to that particular type.

The first step in the reconstruction procedure is the construction of the \texttt{MissingETAssociation} objects --- one for each jet, plus the miscellaneous association.
This is performed by a \texttt{METJetAssocTool}, which sets the \texttt{jetLink} and \texttt{isMisc} variables for each association.
Additionally, this tool initializes a map from jet constituents (represented in \texttt{ElementLink} form) to association indices.
This map is kept as a member of the \texttt{MissingETAssociationMap} and is used for search functionality in the rest of the reconstruction procedure.
It acts as a transient cache which is not written to disk in the persistent representation of the \texttt{MissingETAssociationMap}, and is used to keep track of which objects are selected for the \met\ calculation.

After creation, the associations are sequentially filled with information corresponding to each type of hard object that can overlap with the jets (and each other): muons, electrons, photons, and taus.
In each case, the corresponding \texttt{MissingETAssociator} is used to associate jet constituents (or tracks) with the hard objects and then fill the \texttt{MissingETAssociationMap} with the relevant information.
After determining the associations for each hard object (as decribed below), the tool iterates over the associated constituents.
For each constituent/track, the corresponding jet is used to select the correct \texttt{MissingETAssociation}.
In the case where the constituent/track is not associated with any jet, the miscellaneous association is used.
The member variables of this \texttt{MissingETAssociation} are filled or updated with the constituent/track information described in Section~\ref{sec:metassoc}.
This is repeated until all of the constituents/tracks associated with all hard objects have been allocated to an \texttt{MissingETAssociation}.

The exact methods used for associating jet constituents and tracks to leptons and photons vary depending on the object type and are detailed below.
In the cases of tracks and charged PFOs (which are constructed from tracks), the track in question must always pass quality requirements and be associated with the primary vertex to be considered.
Wherever a jet constituent is a charged PFO, its track is used to determine its associations.
In some instances, precise criteria are not listed for some aspects of these methods; in these cases the definitions are chosen or tuned by the user.

Muon reconstruction includes association of tracks with the muon.
The same association is used for \met\ reconstruction.
Only ID tracks are considered, such that ``standalone'' muons reconstructed using only the MS have no associated tracks.
For the purposes of \met\ reconstruction, only the original track is used; any refitting from the muon reconstruction procedure is ignored.

Although muons are minimally ionising particles, the energy they deposit in the calorimeters is not entirely negligible, and may lead to measurable calorimeter signals exceeding noise thresholds.
The calorimeter cells crossed by a muon track can be identified, in particular any cells in topoclusters or neutral PFOs that might contribute to the reconstruction of jets or other objects.
Association of these cells with the muon can be used to avoid double-counting their energy in corrections to the muon momentum and overlapping jets or the \met\ soft term.

Electron, photon and hadronic tau candidates are reconstructed using a combination of topoclusters and ID tracks.
The associations between the particle candidates and their basic constituents, defined using angular proximity or other more complex selection criteria, are recorded prior to further manipulations refining the particle energy/momentum reconstruction, and can be recalled for the purposes of \met\ reconstruction.
Were these associations not retained in the particle schema, they would need to be reproduced by repeating the corresponding matching procedures.
A custom association procedure could also be followed for particle candidates built from basic constituents that do not directly map onto the jet/\met\ constituents.

\subsection{The core soft term}

The core soft term represents the contribution of all constituents/tracks that are not associated with any jet or other hard object.
This is constructed by a tool called a \texttt{METSoftAssociator}.
This functions by iterating over all constituents/tracks and searching the map for the \texttt{MissingETAssociation} in which it is represented.
There are three possible outcomes for each constituent/track:
\begin{enumerate}
	\item It is found in a \texttt{MissingETAssociation} corresponding to a jet, meaning it is associated with a jet and potentially with leptons/photons.
	\item It is found in the miscellaneous association, meaning it is associated with one or more leptons/photons but not with a jet.
	\item Is is not found in any association, meaning it is not associated with any hard object at all.
\end{enumerate}
All constituents/tracks which are not represented in any association (i.e. outcome 3) are included in the core soft term.
This simply consists of a \texttt{MissingET} object containing the vector and scalar sums of the transverse momenta of these objects.

The core soft term is distinguished from a general soft term because the latter is defined only at the analysis level, where different selections on leptons and photons may be applied.
For example, if a downstream selection removes an isolated lepton, its contribution to the \met\ will enter the soft term, despite not being included in the core soft term.
The information necessary for this later redefinition of leptons/photons at analysis level is encoded in the miscellaneous association in this case.

\subsection{Calculating the missing transverse momentum}
\label{sec:calculatingmet}

The final step of the reconstruction procedure is the actual computation of the missing transverse momentum from the association map, core soft term, and hard objects in the event.
The format of this output is a \texttt{MissingETContainer}, consisting of one \texttt{MissingET} for each term and one for the total sum.
This is created by initializing an empty \texttt{MissingETContainer} and sequentially adding each set of hard objects to it via a tool called \texttt{METMaker}.
The user is free to do this in any order they choose (or omit some objects if necessary), but the standard convention is electrons, photons, taus, muons, and finally, jets.
\texttt{METMaker} can be used directly in analysis to reconstruct the \met\ using customized object selections.
Its main functionality is implemented in the function \texttt{rebuildMET(...)}, which takes as input the collection of hard objects to be added to the calculation.
This function creates a new \texttt{MissingET} representing the corresponding term, computes its 2-dimensional momentum vector (excluding objects which fail overlap removal), and inserts it into the \texttt{MissingETContainer}.
In the case of jets, a specialized function \texttt{rebuildJetMET(...)} is used instead.
This constructs the jet and soft terms, ensuring that no overlapping momentum contributions are double-counted by checking against the objects that were included in the earlier terms.
In this way, a consistent and correct calculation is achieved even if the user applied additional selection requirements to the other hard objects before adding them.
The calculation is then completed using the \texttt{buildMETSum(...)} function of \texttt{METMaker}, which adds all of the terms together in a vector sum to compute the total \met, which is also inserted into the \texttt{MissingETContainer}.

\subsubsection{Overlap removal}
\label{sec:objremoval}

Each time \texttt{rebuildMET(...)} or \texttt{rebuildJetMET(...)} is called, it is necessary to add only momentum contributions which are not otherwise included as part of a higher-priority object.
The definition of which objects receive priority can vary based on user choices; a typical example of usage by ATLAS is given in Ref.~\cite{PERF-2016-07}.
In general, when two hard objects overlap, one is removed entirely from the calculation.
If the removed object contains signals that are not part of any retained object, these signals will then be captured in the soft term.

This method requires keeping a record of which signals are forming part of a hard object term, and which are not (and should therefore be included in the soft term).
This is implemented in the form of a transient mutable bitmask, \texttt{useObjectFlags}, which is index-parallel with \texttt{calkey}/\texttt{trkkey}.
The value of each bit encodes whether the corresponding object has been selected for one of the hard object terms.
When a collection of hard objects is added via \texttt{METMaker}, this is checked to determine if the given consituent basis group should be omitted to avoid double-counting it.
Subsequently, \texttt{useObjectFlags} is updated accordingly if any further objects are selected in each step.
To satisfy thread safety requirements discussed in Section~\ref{sec:evolution}, this bitmask and procedure are implemented within a transient helper class \texttt{MissingETAssociationHelper}.
Earlier versions which did not require thread safety (e.g. during LHC Run 2) instead implemented this directly within \texttt{MissingETAssociationMap}.

\section{Analysis interface}
\label{sec:interface}

This section discusses the interface used for applying analysis-specific (re-) calculations of \met, which include:
\begin{itemize}
\item Using analysis-specific object overlap and overlap removal procedures,
\item Applying updated calibrations for selected hard objects,
\item Propagating the impact of systematic uncertainties impacting hard objects through the \met\ calculation,
\item Choosing objects used for calculating the soft \met\ term (cluster-based or track-based), and
\item Applying additional systematic uncertainties to the soft \met\ term.
\end{itemize}

\subsection{(Re-)Calculating missing transverse momentum in analyses}

The first three above operations can be performed using the same workflow for the initial \met\ reconstruction described in Section~\ref{sec:calculatingmet}.
That is to say the functions \texttt{rebuildMET(...)} or \texttt{rebuildJetMET(...)} can be called using analysis-specific selected and calibrated objects.
This permits users to modify their object definitions/calibrations used in the \met\ calculation and even fix any potential bugs without requiring any large-scale reprocessing of data.
As before, the order in which the \met\ terms are rebuilt using \texttt{rebuildMET(...)} defines a priority list, and objects overlapping with prior objects are omitted from inclusion in the term being calculated, whilst the energy/momentum associated with selected object is subtracted from jets that contain them.
At analysis level, \texttt{rebuildJetMET(...)} is called with all the (calibrated) jets in the event.
However, \texttt{METMaker} can be configured using the argument of this function such that different centrally defined ``working points'' for jets included in the \met\ calculation can be applied.
These working points impose additional selection criteria on the jets to reduce the impact of pileup on \met\ reconstruction.
These can impact performance and the resulting systematic uncertainties in a non-trivial way so optimisation of the \met\ working points are also part of the analysis design.
In addition to selecting the working point for jet selection, \texttt{rebuildJetMET(...)} also allows the user to choose the calculation used for the soft term.
Since Run 2, the default soft term calculation is the ``track soft term (TST)'', where only tracks associated to the primary vertex but not associated with prior hard objects are included in the soft term, however the calorimeter-based ``cluster soft term (CST)'' can also be used to account for soft neutral objects at the expense of increased sensitivity to pileup.
Once these functions have been called the total \met\ can be calculated using the \texttt{buildMETSum(...)} function of \texttt{METMaker}.
This procedure can be repeated to calculate the \met\ associated with systematic variations that impact the 4-momenta of the calibrated objects.

The implementation of \texttt{rebuildMET(...)} and \texttt{rebuildJetMET(...)} can include post-hoc corrections to incorrect reconstruction logic in the production of the \texttt{METAssociationMap}, which would typically come in the form of incomplete overlap removal.
When such corrections can be made, this is a major advantage, due to the computational cost and the lead times needed to rerun a corrected reconstruction campaign.
Bugfixes can be provided in this way to analysis users effectively as soon as they are devised and validated.
The scope for analysis-level corrections is significantly broader in the dynamic \met\ EDM.

When calculating the \met\ for new calibrations or systematic variations of hard objects such as electrons and muons, an object called \texttt{ShallowAuxContainer} is used. 
This ``shallow copy'' references back to the original auxiliary container it was created from for reading data members and only stores locally variables that are set explicitly. 
This allows, for example, the 4-vector of an object to be updated by a new calibration, without making a ``deep copy'' of the original object (i.e. copying \textit{all} of its associated variables) which would be more memory and CPU intensive. 
Since the \texttt{METAssociationMap} identifies these objects by reference to their container and index, a method is needed to match the calibrated object back to the original one for the \met\ re-calculation.
This is achieved by decorating the copied object with an \texttt{ElementLink} to the original object.

As well as enabling the propagation of systematic uncertainties impacting hard objects through the \met\ reconstruction, the software also enables the evaluation of the impact of systematic uncertainties on the soft \met\ term.
These uncertainties are applied as variations on the corresponding \met\ object itself.
They are implemented in a \texttt{METSystematicsTool}, which provides a function \texttt{applyCorrection(...)} taking as input the \texttt{MissingET} to be varied (usually a shallow copy of the original) and the corresponding \texttt{MissingETAssociationMap}.
For each component of the systematic uncertainty, this function can be called to modify or vary the \met object accordingly.
In practice, these generally take the form of adjustments to the scale or resolution of the soft term, decomposed into orthogonal components parallel or perpendicular to the total transverse momentum of all hard objects.
In the case of a ``resolution variation'', the soft term component is smeared by a factor randomly sampled from a Gaussian distribution with the appropriate width.
Before \texttt{applyCorrection(...)} is called, the systematics tool must be ``told'' what kind of variation to apply, using another function \texttt{applySystematicVariation(...)}, which takes as an argument an ATLAS-common object specifying a systematic uncertainty definition.
By iterating through all desired uncertainty components affecting the soft term, one can obtain a set of varied \met\ objects encapsulating their effects, which may then be used as inputs to statistical interpretations.
The generality of this interface allows any form of variation to be applied to the soft term in principle.

\subsection{Specialized functionality}

Occasionally it is beneficial for a physics analysis project to construct a customized variant of the \met.
The flexibility of the \met\ reconstruction design makes it readily adaptable to these analysis-specific techniques that may not follow a standard prescription.
For example, the ``recoil'' variable used in \(W\) boson mass measurements~\cite{STDM-2014-18}, which does not involve any jet definition, can be reconstructed with these tools.
Two further examples are given below in more detail to demonstrate this principle: \met\ calculated purely from tracks, and \met\ computed as if some subset of the objects in the event are ``invisible'' to the detector.
Note also that in general there are no restrictions on the objects used in the calculation, provided that the association map is constructed using the appropriate constitutents.
For example, the user may choose a different vertex or track selection when building the map, and this is supported transparently.
Typical use cases include analyses using only photons in their final state, which cannot rely on tracking to identify a primary vertex (e.g. Refs.~\cite{HIGG-2019-13,HIGG-2019-02}).

In addition to the choice of track-based or cluster-based soft term calculations, the analysis interface also enables the calculation of an entirely track-based \met to further reduce pileup contamination, at the expense of excluding neutral particles from the calculation (which can degrade its accuracy) and limiting the acceptance of the calculation to that of the tracker.
This calculation is handled by the \texttt{rebuildTrackMET(...)} function of \texttt{METMaker} which takes the same arguments as the \texttt{rebuildJetMET(...)} function.
This functions identically (and calls \texttt{rebuildJetMET(...)} itself), except all tracks associated with jets are counted in the ``soft track term''.
When these tracks are added to the object ordinarily called the `` track soft term'',  the result is in fact the full track-based \met, as it then includes all tracks in the event which pass the quality requirements are associated with the primary vertex.
No other objects such as leptons or photons are included in the calculation, so no overlap removal is required in this case.
For a typical use case of this \met definition, see Ref.~\cite{HIGG-2013-13}.

\texttt{METMaker} also includes functionality to mark a container of objects as ``invisible'', excluding these objects and their overlaps from the \met\ calculation.
The implementation is provided by a function \texttt{markInvisible(...)} in \texttt{METMaker}, which replaces the corresponding call to \texttt{rebuildMET(...)}.
This functions much like \texttt{rebuildMET(...)} in that it creates a \met\ term corresponding to the given collection of objects, accounting for overlaps.
The difference is that it sets that term's \texttt{source} tag to a designated values that indicates that the term is to be ignored in the final, overall \met\ calculation.
When \texttt{buildMETSum(...)}, it will skip any terms which are marked as having an invisible source.
The result is ``what the detector would have seen'' if the given set of particles did not interact with it at all.
This has use cases such as marking muons as invisible in \(Z \to \mu\mu\) events to obtain a sample representative of $Z \to \nu\nu$ events, e.g. for estimating backgrounds in analyses using \met without relying on simulations.

\section{Computational performance}
\label{sec:performance}

In an increasingly resource-limited computing environment, the compact energy overlap representation and dynamic \met\ computation permit significant gains in the CPU and disk cost of providing optimised \met\ quantities for analysis.
These savings are achieved primarily by eliminating redundant operations and information that would otherwise be needed to adapt the computation to diverse physics object selections.

\subsection{Algorithmic efficiency}

The CPU cost of the initial reconstruction step is dominated by the cost of associating \met\ and jet constituents (clusters, tracks or particle flow objects) to the physics objects.
At worst, this scales as the product of the number of physics objects ($\mathcal{O}(10)$) and signal constituents ($\mathcal{O}(10)$), but can be accelerated by using predefined links when the physics objects are built from common constituents, or by preemptively segmenting the object collections into bins in $\eta$ and $\phi$ to minimise the number of required comparisons.

For a benchmark sample of 2018 data, with 50 interactions per event on average, reconstruction of raw data into the basic `AOD' analysis data format in a 8-threaded job takes approximately 2 seconds per event~\cite{ATL-SOFT-PUB-2021-002}, evaluated on a 16-core Intel\textregistered{}Xeon\textregistered{}CPU E5-2630 v3 at 2.40 GHz.
In such a job, each instance of association map construction contributes less than 1\% of the CPU cost. Three instances are run, producing output maps serving consumers of three jet collections\footnote{The three jet collections used here were $R=0.4$ jets built from topoclusters, $R=0.4$ jets built from PFOs, and $R=1.0$ jets built from topoclusters, all using anti-$k_t$, where $R$ is the clustering distance parameter.}, for which the constituent associations and overlap calculations need to be repeated.
This time is comparable within 10\% to the time taken for computing a single \met\ collection in a static \met\ event data model, as the required matching operations are for the most part identical.
Any differences due to applying a stricter object selection in the static model are outweighed by the necessity of rerunning the associations for any change of the object selection.
For comparison, running one jet clustering algorithm takes 1.5--2 times this per event, while building simple sums over calorimeter cells is up to 6 times more expensive due to requiring iteration over nearly 200,000 cells.

At analysis time, the additional computation needed to determine the final \met\ two-vectors is $\mathcal{O}$(1~ms), and essentially negligible, even when the operation needs to be redone for every systematic variation.
In a static EDM, similar operations need to be carried out in order to update object calibrations and apply systematic variations.

\subsection{Size on disk}

The output size for each of the \met\ collections is approximately 1.5~kB per event, two thirds of which is in the \met\ association maps, and the remainder constituted by the core soft term.
Simulated events typically have a larger disk footprint, as large samples are produced of processes of interest, which are often biased towards higher centre-of-mass energies. Using top-quark pair events as a benchmark, the size of the \met\ collections is 3~kB per event, with the same breakdown between the association maps and core soft term container.
The size and scaling of the core soft term content with the event activity may come as a surprise; this is because links to the soft signals contributing to the core soft terms are saved as decorations on the objects, permitting traceability and specialised studies of the soft term composition.
For physics analysis purposes, this additional information may be dropped, saving an additional 0.5~kB per event, an important reduction for end-user analysis formats that are kept as lightweight as possible.

For LHC Run 3 (i.e. since 2022), the ATLAS analysis model \cite{AMGRun3} includes two data formats to cover the needs of most analysis use cases:
\begin{itemize}
\item \texttt{DAOD\_PHYS} contains uncalibrated physics object containers holding all the content needed for analysis-ready calibrations to be applied, with a target event size of 50~kB/event. This is representative of the analysis formats used in Run 2.
\item \texttt{DAOD\_PHYSLITE} is a more streamlined format in which calibrations are applied in advance, minimising the disk footprint as well as the analysis-time CPU, with a target event size of 10~kB/event.
\end{itemize}
In \texttt{DAOD\_PHYS}, two \met\ collections are retained, each making up 3\% of the total event size.
In \texttt{DAOD\_PHYSLITE}, the truncated physics object collections require a recomputation of the \met\ associations, but this in fact permits two further size reductions.
Firstly, only one standardized jet collection is retained, hence only one \met\ collection need be saved.
Secondly, as fewer object overlaps need to be registered, the size of the \met\ association maps is also substantially reduced, resulting in a total size of 0.3~kB/event.
For comparison, the total \met\ content in Run 1 consolidated analysis formats was 10\% of the event content, which contained 27 custom \met\ definitions, in a file format that was also less optimised for efficient storage.

Overall, substantial savings are achieved in the computational resources needed to reconstruct \met\, with the added benefit of an increased flexibility for optimisation and refinements of the \met\ reconstruction strategy.

\section{Adaptability for Run 3 \& beyond}
\label{sec:evolution}

The \met\ EDM and reconstruction algorithms described above were originally developed and used at ATLAS for Run 2 of the LHC.
However, the flexibility of this design makes it easily adaptable to the evolving needs of physicists in Run 3 and beyond into the era of the High Luminosity LHC.
This section discusses two particular elements which have been developed for usage in this scope: support for multithreaded processing, and global particle flow reconstruction.

\subsection{Multithreading}

In order to best make use of limited hardware resources in the face of ever-increasing computational demands, all ATLAS reconstruction software is designed for multithreaded use from Run 3 onwards~\cite{athenaMT}.
This introduces several complications into the design of all EDM and reconstruction software, including for \met.
One requirement for the thread-safe design is that all of the \met\ EDM classes must not have their states modified after they have been recorded to the event store (otherwise, multiple threads may attempt to do this simultaneously, thus spoiling the information used by the other).
Effectively, this means no mutable data members are permitted.
In order to meet this requirement, the purely transient \texttt{MissingETAssociationHelper} class was introduced to handle the object selection flags discussed in Section~\ref{sec:objremoval}, which would otherwise require mutable members on the \texttt{MissingETAssociationMap}.

When the association map has been created and \met\ needs to be computed, the user initializes a \texttt{MissingETAssociationHelper}.
As the only argument to its constructor, the user provides a pointer to the relevant \texttt{MissingETAssociationMap}.
The \texttt{MissingETAssociationHelper} then internally initializes one bitmask per association, which can be freely modified since it is thread-local.
After this, the user no longer needs to interact directly with the association map: the helper itself can be provided to \texttt{METMaker} instead.
As each object is sequentially added to the calculation, the helper's object selection flags are dynamically updated, ensuring a correct and consistent application of the overlap removal.
This allows the state of the association map itself to remain unchanged throughout the procedure.

\subsection{Global particle flow}

In addition to multithreading, the \met\ reconstruction software is also extended to support ``global particle flow'' in Run 3.
This is a method for considering a global event description, where each PFO is treated as a particle candidate~\cite{CMS-PRF-14-001}.
It entails uniquely associating PFOs with reconstructed objects, including photons, electrons, muons, and taus, based on the objects that were used to reconstruct them.
It aims to provide a more accurate ambiguity resolution than could be achieved by retrospectively determining associations downstream, thereby allowing more refined calibrations and improving the resolution of \met\ reconstruction.
For example, PFOs that are associated with (or `labelled as') electrons can be excluded from calibrations accounting for charged hadrons.

Photons, electrons, muons, and taus are reconstructed from the same tracks that are used to build PFOs.
Therefore, the association for these is straightforward: a charged PFO is associated with any of these objects if its track was also used to reconstruct that object.
The situation is somewhat more complicated when associating these objects with neutral PFOs, since all use calorimeter cells in their reconstruction but not precisely the same topoclusters that are used to construct PFOs.
However, electron, photons, and taus make use of the same topoclusters during their reconstruction, so the global particle flow method allows a direct assocation regardless.
Associating neutral PFOs with muons is done by identifying the calorimeter cells crossed by the muon's track, as described in Section~\ref{sec:recomap}.

After the PFOs and other objects are initially reconstructed, these associations are determined and saved to each object as a dynamic decoration.
This takes the form of an \texttt{ElementLink} which points to the associated PFO.
Links pointing from the PFO to the associated objects are created similarly.
When the \met\ association map is then created, these links can simply be dereferenced to find the correct association.
This allows reconstruction of the association map even from data formats which do not contain the detailed information necessary to initally determine the association.

\section{Conclusion}
\label{sec:conclusion}

This paper has presented the design and implementation of the event data model (EDM) used by the ATLAS Collaboration for reconstructing missing transverse momentum, \met, since its second data taking run (Run 2).
While defined for the particular purposes of ATLAS, the principles underlying this design generalise to other collider experiments.

This improved EDM enables the flexible recalculation of \met at analysis level using analysis-specific physics object selections and choices of overlap removal.
It also allows the \met to be recalculated using updated calibrations for selected ``hard'' objects, and enables systematic uncertainties associated with these objects to be propagated through the \met calculation.
It supports analysis-specific choices for calculating the soft \met term (e.g. cluster-based or track-based) and applying systematic uncertainties associated with this calculation.
This design also results in more efficient usage of computing resources, as the most performance-intensive part (associating constituents with hard objects) only needs to be performed once, even if multiple object definitions are used or subsequently changed.
The structure of the EDM is also space-efficient, allowing full information for customizable \met calculation to be retained in even the most streamlined analysis data formats.
The flexible and modular design also makes the EDM and reconstruction algorithms easily adaptable to the evolving needs of Run 3 and the future, facilitating fully multithreaded computing and adaptation or replacement of any part of the methodology to suit changes in requirements.

\backmatter


\section*{Acknowledgments}

This work was done as part of the offline software research and development programme of the ATLAS Collaboration, and we thank the collaboration for its support and cooperation.
This project has received funding from the European Research Council (ERC) under the European Union's Horizon 2020 research and innovation programme (grant agreement no. 787331).

\bibliography{met-sw-paper}

\end{document}